\documentclass[prd,superscriptaddress,nofootinbib,notitlepage]{revtex4-1}
\usepackage{epsfig,amsmath,amssymb,slashed}
\usepackage{graphicx}
\usepackage{color}
\usepackage{siunitx}
\usepackage{color}
\usepackage{bm}
\usepackage{subfig}

\newcommand{\ket}[1]{|#1\rangle}

\newcommand{\vacm}{| 0_{\rm M} \rangle}
\newcommand{\vacmb}{\langle 0_{\rm M} |}
\newcommand{\di}{{\rm d}}
\newcommand{\ii}{i}

\def\wT{{\widehat T}}
\def\wj{{\widehat j}}
\def\wJ{{\widehat J}}
\def\wK{{\widehat K}}
\def\wP{{\widehat P}}

\def\wQ{{\widehat Q}}

\def\wpsi{{\widehat{\psi}}}
\def\wrho{{\widehat{\rho}}}
\def\wPi{{\widehat\Pi}}

\def\wa{\widehat a}
\def\wad{\widehat a^{\dagger}}
\def\war{\widehat a^{(R)}}
\def\wadr{\widehat a^{\dagger(R)}}
\def\wal{\widehat a^{(L)}}
\def\wadl{\widehat a^{\dagger(L)}}

\newcommand{\tr}{{\rm tr}}  
  
\newcommand{\e}{{\rm e}}

\newcommand{\omegav}{\boldsymbol{\omega}}

\newcommand{\x}{{\rm x}}

\newcommand{\be}{\begin{equation}}
\newcommand{\ee}{\end{equation}}                                                                               
\def\bea{\begin{eqnarray}}
\def\eea{\end{eqnarray}}                                                                               
\begin{document}

\title{Thermodynamic equilibrium with acceleration and the Unruh effect} 

\author{F. Becattini}
\affiliation{Universit\`a di Firenze and INFN Sezione di Firenze, Florence, Italy}

\begin{abstract}
We address the problem of thermodynamic equilibrium with constant acceleration along
the velocity field lines in a quantum relativistic statistical mechanics framework. 
We show that for a free scalar quantum field, after vacuum subtraction, all mean 
values vanish when the local temperature $T$ is as low as the Unruh temperature 
$T_U = A/2\pi$ where $A$ is the magnitude of the acceleration four-vector. We argue 
that the Unruh temperature is an absolute lower bound for the temperature of any 
accelerated fluid at global thermodynamic equilibrium. We discuss the conditions of 
this bound to be applicable in a local thermodynamic equilibrium situation. 
\end{abstract}

\maketitle

\section{Introduction}
\label{intro}

The study of relativistic matter and quantum fields at thermodynamic equilibrium 
under different conditions is drawing great attention lately. There are several 
motivations behind this interest. On one hand, the relativistic stress-energy tensor 
is the key ingredient in general relativity, hence in relativistic astrophysics and 
cosmology, where one would like to calculate its mean value including quantum effects
at local thermodynamic equilibrium. Furthermore, the apparently successful description 
of the Quark-Gluon Plasma formed in nuclear collisions as a relativistic fluid at 
local thermodynamic equilibrium with acceleration and vorticity \cite{star} has 
stimulated new theoretical developments in this direction. On the other hand, 
when temperature becomes very low and we get close to vacuum state, it is known 
that acceleration involves peculiar quantum-relativistic effects, which were first 
pointed out by Unruh \cite{unruh}. Indeed, the Unruh effect and its consequences 
is still a vibrant subject of investigation (see for instance refs.~\cite{arias,
lambiase}). 

In this paper, we will study an accelerated system at any finite temperature in the 
framework of quantum field theory and quantum statistical mechanics.
The equilibrium thermal state will be defined by the proper density operator:
$$
  \wrho = \frac{1}{Z} \exp [-\widehat H/T_0 + a \widehat K_z/T_0]
$$
where $\widehat H$ is the Hamiltonian and $\widehat K_z$ is the boost operator. 
As we will show, this density operator represents a system at global thermodynamic 
equilibrium with non-vanishing acceleration and finite local temperature, whose 
relation with the constants $T_0$ and $a$ will become clear later on.

We will show - for the free scalar field case - that this equilibrium state has 
a remarkable feature, namely its local temperature, measured by a comoving thermometer
cannot be lower than:
$$
  T_U = \frac{|A|}{2\pi}
$$
where $|A|$ is the magnitude of the acceleration four-vector, which can be defined
as the comoving Unruh temperature. We will argue that this feature extends to any 
fluid and it is not in fact limited to free fields.

\subsection*{Notation}

In this paper we use the natural units, with $\hbar=c=K=1$.\\ 
The Minkowskian metric tensor is ${\rm diag}(1,-1,-1,-1)$; for the Levi-Civita
symbol we use the convention $\epsilon^{0123}=1$.\\  
Operators in Hilbert space will be denoted by a large upper hat, e.g. $\wT$ while unit 
vectors with a small upper hat, e.g. $\hat v$. 

\section{Equilibria in relativistic statistical mechanics}
\label{relstatmech}

In thermal quantum field theory the usual task is to calculate mean values of physical 
quantities at thermodynamic equilibrium. The corresponding density operator in flat
spacetime is:
\be\label{homo1}
 \wrho = (1/Z) \exp [- \widehat H/T_0 + \mu_0 \wQ/T_0]
\ee
where $T_0$ is the temperature and $\mu_0$ the chemical potential (the reason for
the 0 superscript will become clear soon) coupled to a conserved charge $\wQ$, and 
$Z$ the partition function. The above density operator can be made manifestly covariant 
by introducing the four-temperature $\beta = (1/T) u$ where $u$ is the four-velocity 
of the comoving observer; thereby, the equation (\ref{homo1}) can be rewritten as: 
\be\label{homo2}
 \wrho = (1/Z) \exp [- \beta \cdot \widehat P + \mu_0 \wQ/T_0]
\ee
where $\widehat P$ is the four-momentum operator. Note that $T = 1/\sqrt{\beta^2}$ 
is a relativistic invariant; it is the temperature measured by a comoving thermometer, 
according to the most widely accepted formulation of relativistic thermodynamics 
\cite{becalocal}.    

However, the density operator (\ref{homo1}), is not the only form of global thermodynamic 
equilibrium, defined, in general, as a state where the entropy $S=-\tr (\wrho \log \wrho)$ 
is maximal - hence constant - with specific constraints. For instance, it is well known 
\cite{landau,vilenkin} that in non-relativistic quantum mechanics the operator:
\be\label{rotating}
  \wrho = (1/Z) \exp [-\widehat H /T_0 + \omega \widehat J_z/T_0 + \mu_0 \wQ/T_0]
\ee
$\widehat H$ the hamiltonian and $\widehat J_z$ the angular momentum operator along 
some axis $z$, represents a globally equilibrated spinning fluid with angular velocity 
$\omega$. 

The above (\ref{homo2}) and (\ref{rotating}) are indeed special cases of the most 
general thermodynamic equilibrium density operator, which can be obtained by maximizing 
the total entropy $S = -\tr (\wrho \log \wrho)$ with the constraints of given 
energy-momentum and charge densities at some specific “time” over some spacelike 
hypersurface $\Sigma$ \cite{weert,becalocal,nippon}. Therefore, the general equilibrium 
density operator can be written in a fully covariant form as \cite{weert,zubarev,
becacov}: 
\be\label{gencov}
  \wrho = (1/Z) \exp \left[- \int_\Sigma \di\Sigma_\mu  \left( \wT^{\mu\nu} \beta_\nu 
- \zeta \wj^\mu \right) \right]
\ee
where $\wT^{\mu\nu}$ is the stress-energy tensor, $\wj^\mu$ a conserved current and 
$\zeta$ is a scalar whose meaning is the ratio between comoving chemical potential 
comoving temperature. The four-vector field $\beta$ has the physical meaning of 
the inverse proper temperature times the four-velocity and, in general, does not 
need to be constant and uniform at equilibrium.

Indeed, for the right hand side of eq. (\ref{gencov}) to be a good equilibrium 
distribution, the integral must be independent of $\Sigma$, which also means independent 
of time if $\Sigma$ is chosen to be a $t=const$ hypersurface. Provided that 
the flux at some timelike boundary vanishes, this condition requires the divergence 
of the vector field in the integrand to be zero and this in turn \cite{becacov} 
that the scalar $\zeta$ is constant and $\beta$ a Killing vector field, that is fulfilling 
the equation \footnote{For general non-cartesian coordinates can be involved, we 
use covariant derivative notation even though we are working in flat spacetime}:
\be\label{kill}
  \nabla_\mu \beta_\nu + \nabla_\nu \beta_\mu = 0  
\ee
The density operator (\ref{gencov}) is also well suited to describe thermodynamic 
equilibrium in a general curved spacetime possessing a timelike Killing vector field. 
In Minkowski spacetime, which we will be dealing with in this work, the general 
solution of the eq.~(\ref{kill}) is:
\be\label{killsol}
   \beta_\mu = b_\mu + \varpi_{\mu\nu} x^\nu
\ee
where $b$ is a constant four-vector and $\varpi$ a constant antisymmetric tensor, 
which, because of eq.~(\ref{killsol}) can be written as an exterior derivative of the
$\beta$ field, that is $\varpi_{\nu\mu} = -\frac{1}{2} (\partial_\nu \beta_\mu - 
\partial_\mu \beta_\nu)$. Hence, by using the eq.~(\ref{killsol}), the integral 
in eq.~(\ref{gencov}) can be rewritten as:
\be\label{generatkill}
  \int_\Sigma \di\Sigma_\mu \; \wT^{\mu\nu} \beta_\nu = - b_\mu {\wP}^\mu  
  + \frac{1}{2} \varpi_{\mu\nu} \wJ^{\mu\nu}
\ee
and the density operator (\ref{gencov}) as:
\be\label{gener2}
  \wrho = \frac{1}{Z} \exp \left[ - b_\mu {\wP}^\mu  
  + \frac{1}{2} \varpi_{\mu\nu} \wJ^{\mu\nu} + \zeta \wQ \right]
\ee
where the $\wJ$'s are the generators of the Lorentz transformations:
\be\label{generators}
 \wJ^{\mu\nu} = \int_{\Sigma} \di \Sigma_\lambda \; \left( 
 x^\mu \wT^{\lambda\nu} - x^\nu \wT^{\lambda\mu} \right) 
\ee
Therefore, besides the chemical potentials, the most general equilibrium density 
operator in Minkowski spacetime can be written as a linear combinations of the 10 
generators of the Poincar\'e group with 10 constant coefficients. 

It can be readily seen that the density operator (\ref{homo2}) is obtained by setting 
$b=(1/T_0)(1,0,0,0)$ and $\varpi=0$, what we define as {\em homogeneous thermodynamic 
equilibrium}. The rotating global equilibrium in eq.~(\ref{rotating}) can be obtained 
as a special case of eq.~(\ref{gener2}) by setting:
\be\label{rot}
 b_\mu = (1/T_0,0,0,0) \qquad \qquad \varpi_{\mu\nu} = (\omega/T_0) (g_{1\mu} g_{2\nu} 
- g_{1\nu} g_{2\mu})
\ee
i.e. by imposing that the antisymmetric tensor $\varpi$ has just a ``magnetic"
part; thereby, $\omega$ gets the physical meaning of a costant angular velocity 
\cite{landau}. However, there is a third, not generally known, form which is conceptually
independent of the above two, which can be obtained by imposing that $\varpi$ has
just an "electric" (or longitudinal) part, i.e.:
\be\label{acc}
 b_\mu = (1/T_0,0,0,0) \qquad \qquad \varpi_{\mu\nu} = (a/T_0) (g_{0\nu} g_{3\mu}
 - g_{3\nu} g_{0\mu})
\ee
The resulting density operator is:
\be\label{accdo}
  \wrho = (1/Z) \exp \left[ -\widehat H /T_0 + a \widehat K_z/T_0 \right]
\ee
$\widehat K_z \equiv \widehat J_{30}$ being the generator of a Lorentz boost along 
the $z$ axis. As we will see, this density operator represents a relativistic fluid
with constant comoving acceleration along 
the $z$ direction and the combination $\widehat H - a \widehat K_z$ can be seen as
the generator of translation along its flow lines \cite{leinaas}. Note that the 
operators $\widehat H$ and $\widehat K_z$ are both conserved and yet, unlike in 
the rotation case (\ref{rotating}) they {\em do not} commute with each other.
In fact, the boost operator $\widehat K_z$ is {\em explicitely} time dependent 
as, from (\ref{generators}):
$$
 \widehat K_z = \widehat J_{30} = t \widehat P_z - \int \di^3 x z \wT^{00}
$$
and its Heisenberg equaton of motion reads:
$$
  \ii \frac{\di \widehat K_z}{\di t} = [\widehat K_z,\widehat H] + \ii \frac{\partial
  \widehat K_z}{\partial t} = -\ii \widehat P_z + \ii \widehat P_z = 0
$$
This makes the density operator (\ref{accdo}) a very peculiar kind of thermodynamic 
equilibrium. 

The question arises whether and how a density operator like (\ref{accdo}) can be 
realized. In quantum statistical mechanics, the density operator (\ref{accdo}) 
is the solution of the maximization of entropy constrained by fixed mean values
of the total energy $\langle \widehat H \rangle = H_0$ and of the mean boost 
$\langle \widehat K_z \rangle = K_0$. In flat space-time, it is possible to shift
the origin of the coordinates to the centre-of-mass according to:
$$
  \widehat K'_z = \widehat K_z - t \widehat P_z + z_{CM} \widehat H
$$
so as to make $K'_0 = 0$. Thereby, a state such as (\ref{accdo}) seems to be irrelevant. 

Nevertheless, for systems at local thermodynamic equilibrium, the operator (\ref{gener2})
is the first order local expansion of the four-temperature field \cite{becalocal}
and must then be a better approximation to describe a fluid with non-vanishing local 
acceleration (and vorticity) with respect to the local four-temperature vector alone.
Similarly, in presence of the gravitational field, the (\ref{accdo}) is the leading 
order expansion of the generally covariant expression (\ref{gencov}) in inertial 
coordinates \cite{becacurv} and it can then describe thermodynamic equilibrium in 
a constant and uniform gravitational field. This is in agreement with the principle
of equivalence and it is confirmed by taking its non-relativistic limit. 
For it is constant, we can write the explicit form of its 
exponent at $t=0$:
$$
  \widehat H - a \widehat K_z = \int \di^3 \x \; (1 + az) \wT^{00}
$$
For the single particle in non-relativistic quantum mechanics, the operator $\wT^{00}$ 
is (restoring $c$) reads:
$$ 
 \wT^{00} = (mc^2 + \widehat p^2/2m) \delta^3({\bf x} - \widehat{\bf x})
$$
where $\widehat{\bf x}$ is the position operator. Hence:
$$
  \widehat H - a \widehat K_z = (mc^2 + \widehat p^2/2m) \int \di^3 \x \; (1 + az/c^2) 
\delta^3({\bf x} - \widehat{\bf x}) \simeq mc^2 + \widehat p^2/2m + ma \widehat z
$$
which is the hamiltonian of a non-relativistic particle in a constant gravitational 
field $a$; this demonstrates our interpretation. Of course, for real gravitational
fields, the (\ref{accdo}) is an approximation of (\ref{gencov}) without curvature
terms, which may play a role at extremely low temperatures, when the curvature
scale becomes comparable with thermal wavelengths. 

The mixed states (\ref{rotating}) and (\ref{accdo}) are the two main thermodynamic
equilibria with $\varpi \ne 0$ in eq.~(\ref{killsol}) in flat spacetime; all other 
cases are combinations thereof. It is important to stress that in 
both cases the four-temperature vector $\beta$ is not a global timelike Killing vector. 
In fact, there is a non-trivial Killing horizon hypersurface defined by $\beta^2=0$ 
dividing the spacetime into regions where $\beta$ is timelike and spacelike respectively. 
This does not, though, hamper the calculation of mean values of local operators 
in the regions where $\beta$ is timelike and future-oriented, as we will see in 
detail in the next section.

\section{Equilibrium with acceleration}
\label{acceq}

The physics described by an equilibrium density operator is contained in the four-temperature 
$\beta$. As has been mentioned in the Introduction, whenever $\beta$ is timelike 
and future-oriented, its magnitude is the inverse comoving temperature, whilst its 
direction is a flow velocity. In the rotating case (\ref{rot}) 
one has:
$$
\beta = \frac{1}{T_0}(1,\omegav \times {\bf x})
$$
and the velocity field is that of a rigid rotation \cite{landau} up to the radius $r$
where $\omega r =1$. The inverse of the
time component $\beta^0$ is the temperature measured by a thermometer at rest with 
the inertial observer, while $1/\sqrt{\beta^2}$ is the {\em proper} temperature $T$
measured by a comoving thermometer. Thus, in the rotating case the temperature measured 
by the inertial observer who sees the fluid in a rotational motion is uniform, constant
and equal to $T_0$ while latter is constant but not uniform, and it is related to 
$T_0$ by the so-called Tolman's law $T=T_0/\sqrt{1-|\omegav \times {\bf x}|^2}$. 

\begin{figure}[ht]
\includegraphics[width=0.8\columnwidth]{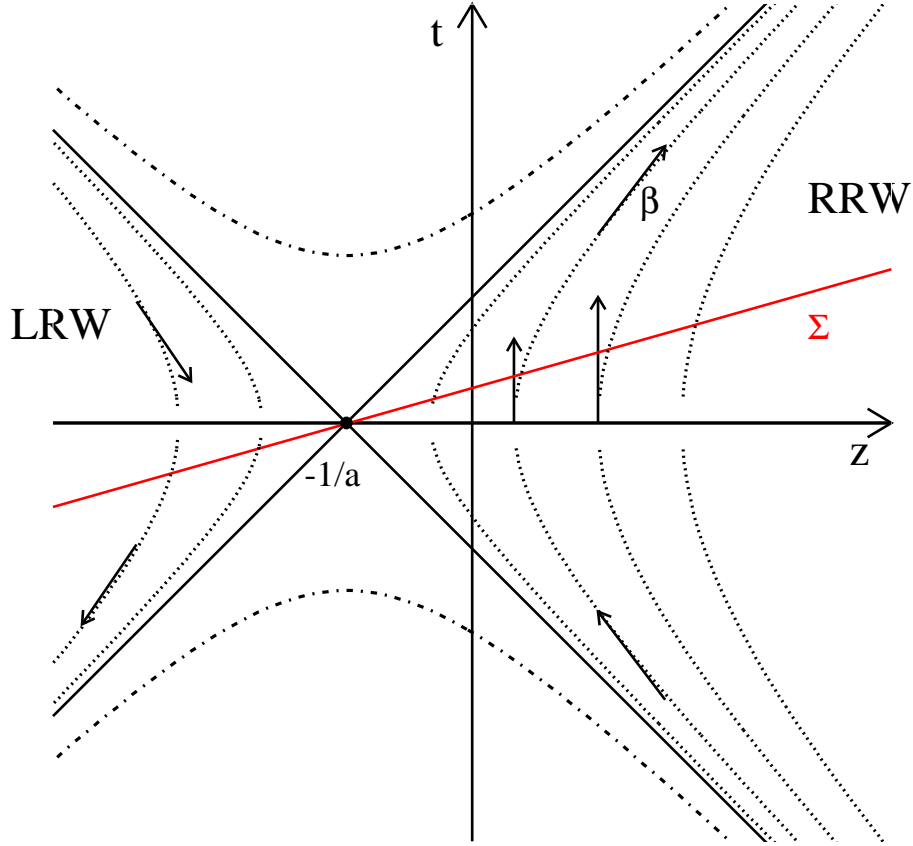}
\caption{(Color online) 2D Minkowski space-time diagram with the field lines (dashed)
of the four-temperature $\beta$ Killing field (\ref{betacc}) in the Right Rindler Wedge 
(RRW) and the Left Rindler Wedge (LRW). In these wedges the vector field (drawn 
with arrows) is time-like and it is future-directed only in the RRW. Also shown a 
hyperplane $\Sigma$ perpendicular to $\beta$ which can be used as space-like hypersurface 
to define the Klein-Gordon inner product and calculate the operators $\wPi_R$,$\wPi_L$. 
The quantum field in the RRW is causally disconnected from the LRW.}
\label{figura}
\end{figure}

In the accelerating case (\ref{acc}), the contravariant components of $\beta$ read: 
\be\label{betacc}
 \beta^\mu= \frac{1}{T_0}\left(1 + a z, 0, 0,a t \right)
\ee
To understand the meaning of this vector field, it is very useful to shift the origin 
of the coordinates in $z=-1/a$. Thereby, the four-temperature becomes, with $z'=
z + 1/a$:
\be\label{betacc2}
 \beta^\mu = \frac{a}{T_0}\left(z', 0, 0,t \right)
\ee
The field lines are hyperbolae with constant values of $k=\sqrt{z'^2 - t^2}$ and 
the four-temperature is timelike only outside the light cone of $(0,0,0,-1/a)$ in the 
so-called right and left Rindler wedges (RRW and LRW respectively), see fig.~\ref{figura}. 
Besides, the four-temperature is future oriented only in the RRW. In this region, 
the comoving temperature reads:
\be\label{tt0}
  T = \frac{1}{\sqrt{\beta^2}} = \frac{T_0}{ka}
\ee
and it is constant along the flow lines. The eq.~(\ref{tt0}) can be seen as an instance
of Tolman's law \cite{buchholz}, but it is most naturally obtained within this 
approach by imposing the four-temperature to be a Killing vector. Indeed, the constancy
along the flow lines is a general property of Killing vectors; by double-contracting 
the equation (\ref{kill}) with $\beta$:
\be
  0 = \beta^\mu\beta^\nu (\nabla_\mu \beta_\nu + \nabla_\nu \beta_\mu) = 
  2 \beta^\mu\beta^\nu \nabla_\mu \beta_\nu = \beta^\mu \nabla_\mu \beta^2 \equiv 
  \sqrt{\beta^2} D \beta^2 
\ee
Indeed, it is an expected feature of thermodynamic equilibrium that observers moving 
along the velocity field $u \propto \beta$ see no change in temperature and, consequently, 
in all other thermodynamic quantities \cite{kovtun,becalie}. 

Because of (\ref{betacc2}) and (\ref{tt0}), the velocity field $u= \beta/\sqrt{\beta^2}$
reads:
$$
 u^\mu = \frac{1}{k} (z', 0, 0, t)
$$
and its derivative along the flow $u$ - that is the acceleration $A^\mu$ - reads:
$$
  A^\mu = \frac{1}{k^2} (t, 0, 0, z')
$$
implying $A^2 = - 1/k^2$, i.e. $A^2$ is constant along the flow lines. The motion
described by the two above relations is a well known one in special relativity, 
the so-called uniformly accelerated observer. Note that:
\be\label{comovt}
   - \frac{A^2}{T^2} = \frac{a^2}{T_0^2}
\ee
implying that $a/T_0$ is the constant ratio between the magnitude of the acceleration
four-vector $A$ and that of the proper temperature measured by the comoving observer 
along the flow line. Finally, the temperature measured by the inertial observer 
in the RRW is the inverse of the time component of $\beta$ in eq.~(\ref{betacc2}):
$$
  T_{\rm inertial} = T_0 \frac{1}{az'} 
$$
and, along the flow line:
$$
  T_{\rm inertial} = T_0 \frac{1}{a \sqrt{k^2+t^2}}
$$
that it, it {\em decreases} for far times, when the speed of the system approaches
$c$.

The Killing vector field (\ref{betacc}) has normal hypersurfaces, which are the 
hyperplanes (see fig.~\ref{figura}):
$$
  t = k (z + \frac{1}{a}) = k z'
$$
The existence of normal hypersurfaces is allowed by the vanishing vorticity of
the field (\ref{betacc}). Indeed, as it can be readily checked:
$$
  \epsilon^{\mu\nu\rho\sigma} \beta_\sigma \partial_\nu \beta_\rho = 0
$$
%

\section{Factorization of the equilibrium density operator}
\label{operator}

We now come to the study of the equilibrium density operator (\ref{accdo}). Firstly, 
we note that the operator in the exponent is a Lorentz boost with respect to the displaced 
origin $(0,0,0,-1/a)$, what follows from the general formula:
$$
  \wJ'^{\mu\nu} \equiv \wJ^{\mu\nu}_x = \wJ^{\mu\nu} - \left( x^\mu \wP^\nu - 
  x^\nu \wP^\mu \right)
$$
implying that:
\be\label{kshift}
   \wK^\prime_z = \wK_z - \frac{1}{a} \widehat H
\ee
The density operator (\ref{accdo}) can then be rewritten as: 
\be\label{accdo2}
  \wrho = \frac{1}{Z} \exp \left[ a \wK^\prime_z/T_0 \right]
\ee

As we discussed in Section \ref{relstatmech}, this is a special case of the generally covariant 
form (\ref{gencov}) with $\beta$ given by the equations (\ref{betacc}) or (\ref{betacc2}), 
with an arbitrary choice of the space-like hypersurface $\Sigma$, e.g. the horizontal 
$t=0$ hyperplane. The nice feature of the density operator in eq.~(\ref{gencov}) 
is that the Killing vector $\beta$ (\ref{betacc2}) vanishes in one point, which 
is the displaced origin $(0,0,0,-1/a)$. This is at variance with, e.g. the rotating 
case in (\ref{rot}) and it implies that {\em any} spacelike hypersurface through 
this point will allow to separate the exponent of (\ref{accdo}) into two commutating - 
hence factorizable - operators involving the field degrees of freedom on either 
side. The reason is that the only contribution to the commutator:
$$
\left[ \int_{z' > 0} \di \Sigma_\mu \wT^{\mu\nu} \beta_\nu, 
 \int_{z' < 0} \di \Sigma_\mu \wT^{\mu\nu} \beta_\nu \right]
$$
on a spacelike hyperplane $\Sigma$ (see fig.~\ref{figura}) through $z'=0$ stems 
from the point $z'=0$, i.e. the only point where the field operators have non-vanishing 
commutators; since $\beta=0$ therein, the above commutator vanishes. Thus, defining:
\be\label{gammadef}
\gamma = T_0 \beta
\ee
which is still a Killing vector field, we have:
\be\label{pirpil}
  - a \widehat K^\prime _z = \int \di \Sigma_\mu \wT^{\mu\nu} \gamma_\nu \equiv 
  \wPi = \int_{z' > 0} \di \Sigma_\mu 
  \wT^{\mu\nu} \gamma_\nu + \int_{z' < 0} \di \Sigma_\mu \wT^{\mu\nu} \gamma_\nu
  \equiv \wPi_R - \wPi_L
\ee
with:
$$
  [\wPi_R,\wPi_L] = 0
$$
where the subscripts $R$ and $L$ stand for the right and left Rindler wedge respectively. 
We remark that a $-$ sign in the definition (\ref{pirpil}) of $\wPi_L$ appears because 
the unit vector perpendicular to $\Sigma$ in the LRW has the opposite direction 
with respect to $\beta$ or $\gamma$, see fig.~\ref{figura}. Hence, by using eqs.
~(\ref{gammadef}) and (\ref{pirpil}), one can write:
\be\label{relkpi}
   \wK^\prime_z = -\frac{1}{a} \wPi = -\frac{1}{a}(\wPi_R - \wPi_L) 
\ee
and the density operator (\ref{accdo2}) becomes:
\be\label{accdo3}
 \wrho = \frac{1}{Z} \exp[-\wPi_R/T_0] \exp[\wPi_L/T_0]
\ee
where $R$ and $L$ stands for the right and left Rindler wedge respectively (see 
fig.~\ref{figura}). 

The form (\ref{accdo3}) of the density operator makes it apparent that the mean 
value of a local operator at any point $x$ in either the RRW or the LRW only depends 
on the operator $\wPi_R$ or $\wPi_L$ respectively. Indeed, since each $\wPi_{R,L}$
involves only the field operators in its wedge, one can write, in the Hilbert space 
of the field states:
$$
 \wPi_R = \wPi_R \otimes I \qquad \qquad \wPi_L = I \otimes \wPi_L 
$$
Consequently, the partition function is the product of two independent factors:
$$
 Z = \tr(\exp[-(\wPi_R-\wPi_L)/T_0]) = 
  \tr_R (\exp[-\wPi_{R}/T_0]) \tr_L (\exp[\wPi_{L}/T_0])
$$
where the subscripts $R,L$ indicate that the trace is computed on the Hilbert space 
spanned by the field degrees of freedom on the RRW and LRW respectively. So, if 
$\widehat O(x)$ is a local operator with $x$ lying, e.g. on the RRW, its mean value 
will be:
\be\label{meanv}
  \langle \widehat O(x) \rangle \equiv \frac{1}{Z} \tr (\exp[-a \widehat K'_z] 
  \widehat O(x)) = \frac{1}{Z_R} \tr_R (\widehat O(x) \exp[-\wPi_R/T_0])
\ee
regardless of the field states in the LRW.

The operators $\wPi_{R,L}$ are the generators of translations along the hyperbolic 
$\beta$ field lines. This can be readily proved by calculating the commutator of the
combination $\wPi_R - \wPi_L$ with the fields operators. If $x \in $ RRW: 
\begin{align*}
 &  [\wPi_R - \wPi_L,\wpsi(x)] = [\wPi_R,\wpsi(x)] = - T_0  
 \left[ b \cdot \wP - \frac{1}{2} \varpi : \wJ, \wpsi(x) \right] \nonumber \\
 & =- \ii T_0 \left( b \cdot \partial \wpsi - \frac{1}{2} \varpi_{\mu\nu} 
  \left( x^\mu \partial^\nu - x^\nu \partial^\mu \right) \wpsi \right) = 
- \ii T_0 \beta \cdot \partial \wpsi = - \ii \gamma \cdot \nabla \wpsi
\end{align*}
where we have used the eqs.~(\ref{generatkill}),(\ref{gammadef}),(\ref{pirpil}), 
the known commutation relations of the Poincar\'e algebra and the fact that $\wpsi(x)$ 
commutes with $\wPi_L$ if $x \in $ RRW. This happens because $\wPi_L$ is formed 
with field operators which are causally disconnected from those in the the RRW. 
Defining a parameter $\tau$ such that $\gamma^\mu = \di x^\mu/\di \tau$, we have:
\be\label{commfield}
  [\wPi_R,\wpsi(x)] = - \ii \gamma \cdot \nabla \wpsi(x) = 
 - \ii \frac{\partial}{\partial \tau} \wpsi(x)
\ee
i.e. the coordinate $\tau$ plays the role of a time along the $\gamma$ field lines.
Accordingly, for the LRW, one has:
\be\label{commfield2}
  [\wPi_L,\wpsi(x)] = \ii \gamma \cdot \nabla \wpsi(x) = 
  \ii \frac{\partial}{\partial \tau} \wpsi
\ee
%

\section{The free scalar field in Rindler coordinates}

We will now consider in more detail the simplest instance of a quantum field theory: 
the free real scalar field. This a well known problem in Rindler coordinates: we 
refer to the nice review in ref.~\cite{crispino}, here we will just present a short
summary. The first step is to define an inner product of functions:
\be\label{innprod}
  (\phi_1,\phi_2) = i \int_\Sigma \di \Sigma_\mu \; \left( \phi_1^* \nabla^\mu \phi_2
 - \phi_2 \nabla^\mu \phi_1^* \right)
\ee
where $\Sigma$ is a spacelike hypersurface with future-oriented normal vector in 
Minkowski spacetime, e.g. $t=const$. This is also called Klein-Gordon inner product 
because it is independent of $\Sigma$ if $\phi_1,\phi_2$ are solutions of the Klein-Gordon 
equation $\Box \phi + m^2 \phi = 0$. In this case, it is easy to 
realize that the vector field integrand in (\ref{innprod}) is divergenceless, hence 
if the functions $\phi_1,\phi_2$ or their normal derivatives vanish at the timelike 
boundary of a closed spacetime region, the scalar product can be calculated over 
any spacelike hypersurface. It can be checked that: 
$$
  (\phi_1,\phi_2) = (\phi_2,\phi_1)^*=-(\phi_1^*,\phi_2^*)^* \implies (\phi_1^*,\phi_2)
= - (\phi_2^*,\phi_1)
$$
and that the Klein-Gordon inner product (\ref{innprod}) is a sesquilinear form, 
that is antilinear for the first argument. 

In general, the real scalar quantum field $\wpsi(x)$ can be expanded in normalized 
(according to the inner product (\ref{innprod}) solutions of the Klein-Gordon equation:
\be\label{expand}
 \wpsi (x) = \sum_i u_i \wa_i + u^*_i \wad_i 
\ee
The function $u$ is an eigenfunction of the derivative along the timelike Killing 
field (in this case $\beta$ or $\gamma$) and has a positive inner norm, which sets
a distinction between creation and destruction operator in eq.~(\ref{expand}) 
\cite{carroll}:
\be\label{orthon}
 (u_i,u_j) = \delta_{ij} \implies (u^*_i,u^*_j) = -\delta_{ij} \qquad (u_i^*,u_j)
= 0 \implies (u_i,u^*_j) = 0
\ee
The eq.~(\ref{expand}) can be inverted to obtain the operators $\wa,\wad$:
\be\label{aadag}
  \wa_j = (u_j,\wpsi)  \qquad  \wad_j = - (u^*_j,\wpsi)
\ee
By using the (\ref{aadag}) and the canonical commutation relations of the field, 
it can be shown that \cite{crispino}:
\be\label{comm2}
   [\wa_i,\wad_j] = \delta_{ij} \qquad [\wa_i,\wa_j] =[\wad_i,\wad_j] = 0
\ee
which are the usual commutation relations between creation and destruction operators. 

The field equation $(\Box + m^2) \wpsi = 0$ can be solved in Rindler coordinates:
\be\label{rrindler}
   \tau = \frac{1}{2a} \log \frac{z'+t}{z'-t} \qquad \qquad
   \xi = \frac{1}{2a} \log [ a^2 (z'^2-t^2)]
\ee
whose inverse read, in the RRW where $z'>0$:
\be\label{rinverse}
  t = \frac{\e^{a\xi}}{a} \sinh (a\tau) \qquad \qquad z'= \frac{\e^{a\xi}}{a} 
  \cosh (a\tau)  
\ee
while the ``transverse" coordinates $x$ and $y$ are cartesian. In the LRW, instead:
\be\label{linverse}
  t = - \frac{\e^{a\bar\xi}}{a} \sinh (a \tau) \qquad \qquad z'= -\frac{\e^{a\xi}}{a} 
  \cosh (a\tau)  
\ee
while the direct (\ref{rrindler}) are maintained. With these definitions, it turns
out that $\di x^\mu /\di \tau = \gamma$ (see eq.~(\ref{pirpil})) both in the RRW and
the LRW. The general solution of the Klein-Gordon equation depends on three parameters: 
$\omega$, a positive real number and a vector ${\bf k}_T$ which is but the transverse 
momentum of a single mode. The normalized (according to the inner product (\ref{innprod}))
eigenfunction in the RRW reads \cite{crispino}:
\be\label{eigenf}
  u(\tau,\xi,{\bf x}_T)_{\omega {\bf k}_T} = \sqrt{\frac{\sinh(\pi\omega/a)}{4\pi^4 a}} 
  {\rm K}_{\ii \omega/a}\left(\frac{m_T \e^{a\xi}}{a}\right) \e^{\ii {\bf k_T}\cdot
  {\bf x}_T} \e^{-\ii \omega \tau}
\ee
where ${\rm K}$ are modified Bessel functions, and the relevant field expansion 
(\ref{expand}):
\be\label{rfield}
  \wpsi (\tau,\xi,{\bf x}_T)^{(R)} = \int_0^{\infty} \di \omega \int \di^2 {\rm k}_T \; 
  \left( u_{\omega {\bf k}_T} \war_{\omega {\bf k}_T} + u_{\omega {\bf k}_T}^* 
  \wadr_{\omega {\bf k}_T} \right)
\ee
where the creation and destruction operators fulfill the commutation relation 
\be\label{commr}
  [\war_{\omega {\bf k}_T},\wadr_{\omega^\prime {\bf k}^\prime_T}] = 
  \delta(\omega-\omega^\prime) \delta^2({\bf k}_T - {\bf k}^\prime_T)
\ee
Note that from (\ref{eigenf}):
\be\label{tauderiv}
 \ii \frac{\di u}{\di \tau} = \ii \gamma \cdot \nabla u = \omega u 
\ee
that is the $u$'s are eigenfunctions of the transport along the Killing field. 

Similarly, in the LRW, the quantum field can be expanded into eigenfunctions having
the same modes and same functional form of $\tau$ and $\xi$ as the $u$'s in 
eq.~(\ref{eigenf}), with the important proviso that the role of creation and destruction 
operators is interchanged, i.e.:
\be\label{lfield}
  \wpsi (\tau,\xi,{\bf x}_T)^{(L)} = \int_0^{\infty} \di \omega \int \di^2 {\rm k}_T \; 
  \left( u_{\omega {\bf k}_T} \wadl_{\omega {\bf k}_T} + u_{\omega {\bf k}_T}^* 
  \wal_{\omega {\bf k}_T} \right) 
\ee
with, again:
\be\label{comml}
  [\wal_{\omega {\bf k}_T},\wadl_{\omega^\prime {\bf k}^\prime_T}] = 
  \delta(\omega-\omega^\prime) \delta^2({\bf k}_T - {\bf k}^\prime_T)
\ee
The swap between creation and destruction can be understood by recalling that the 
space-like hypersurface chosen to calculate the inner product (\ref{innprod}) has 
the same orientation as the Killing vector $\gamma$ in the RRW, but opposite in 
the LRW. As a consequence, if $u(\tau,\xi,{\bf x}_T)$ is the eigenfunction (\ref{eigenf}) 
in the RRW, one has
\be\label{norm}
 (u_{\omega {\bf k}_T},u_{\omega' {\bf k'}_T}) =
 (u_{\omega {\bf k}_T},u_{\omega' {\bf k'}_T})_{\rm R} = 
 \delta(\omega-\omega^\prime) \delta^2({\bf k}_T - {\bf k}^\prime_T)
\ee
but if it is the eigenfunction with the {\em same} functional dependence on the 
arguments $\tau,\xi,{\bf x}_T$, in the LRW, then 
\be\label{norm2}
 (u_{\omega {\bf k}_T},u_{\omega' {\bf k'}_T}) =
 (u_{\omega {\bf k}_T},u_{\omega' {\bf k'}_T})_{\rm L} = 
 - \delta(\omega-\omega^\prime) \delta^2({\bf k}_T - {\bf k}^\prime_T)
\ee
simply because in the LRW $\di\Sigma_\mu \propto -\gamma_\mu/\sqrt{\gamma^2}$ if $\Sigma$
is the hyperplane orthgonal to $\gamma$ (see fig.~\ref{figura}). 
Looking at the eqs.~(\ref{orthon}), (\ref{aadag}) the swap of creation and destruction 
operators as operatorial coefficients of the functions $u$ and $u^*$ in the LRW 
is apparent. Furthermore, as the fields in the RRW and LRW are causally disconnected, 
one has:
\be\label{discon}
  [\war_{\omega {\bf k}_T},\wal_{\omega' {\bf k'}_T}] = 
  [\war_{\omega {\bf k}_T},\wadl_{\omega' {\bf k'}_T}] = 0
\ee
that is, all commutators of creation and destruction operators of RRW and LRW respectively
vanish.

Finally, the field can be expanded in plane waves, as usual, and equating its expansion 
with the above one in Rindler eigenfunctions one can obtain the Bogoliubov relations 
between the two sets of creation and destruction operators. Defining:
$$
 v_p = \frac{1}{(2\pi)^{3/2}} \frac{1}{\sqrt{2|p^0|}} \exp[-i p \cdot x]
$$
and plugging the plane wave field expansion in the (\ref{aadag}) one obtains, 
for instance:
\be\label{bogo}
  \war = (u_{\omega {\bf k}_T},\wpsi) = (u_{\omega {\bf k}_T},\int \di^3 {\rm p} 
 \; a_p v_p + a^\dagger_p v^*_p ) = \int \di^3 {\rm p} \; (u_{\omega {\bf k}_T},v_p) 
  a_p + (u_{\omega {\bf k}_T},v^*_p) a^\dagger_p
\ee
As it is known, both Klein-Gordon inner products are non-vanishing in eq.~(\ref{bogo})
so that the RRW destruction operators can be obtained as a linear combination of
creation and destruction operators of particles in eigenstates of linear momentum.

\section{Equilibrium mean values}

The calculation of mean values of local observables with the density operator 
(\ref{accdo}) - or its equivalent form (\ref{accdo3}) - is similar to that in the 
familiar thermal field theory. 
Let us consider a local operator which is hermitian and quadratic in the field, such as 
$\wpsi^2$, $\nabla_\mu\wpsi \nabla^\mu \wpsi$, $\nabla^\mu \wpsi \nabla^\nu \wpsi$ etc. 
Most operators of physical interest belong to this class, including the stress-energy 
tensor. By using the expansion (\ref{expand}), and using the subscript $i$ as a 
shorthand for $(\omega,{\bf k}_T)$, its expectation value in the, e.g., RRW can be written 
as:
\be\label{generic}
  \langle A\wpsi B\wpsi \rangle = \sum_{i,j} f_{i,j}(x) \langle \wadr_i \war_j \rangle 
 + f_{i,j}(x)^* \langle \war_i \wadr_j \rangle
\ee
where $A$ and $B$ denote linear operations on the field, like multiplication by a
scalar or derivation and $f_{i,j}$ is a functional expression depending on the specific
$A$ and $B$ and the eigenfunctions $u$; $\langle \; \rangle$ stands for the trace 
on the RRW as in eq.~(\ref{meanv}). An analogous expression will be found in the LRW.

We begin our derivation by calculating the commutation relation between the operators 
$\wPi_{R,L}$ and the creation and destruction operators. As an example, we can derive
$[\wPi_R,\war_i]$ based on the eq.~(\ref{commfield}) and by using eqs.~(\ref{expand}), 
(\ref{orthon}), (\ref{aadag}) and (\ref{tauderiv}): 
$$ 
  [\wPi_R,\war_i] = [\wPi_R,(u_i,\wpsi)_R] = (u_i,[\wPi_R,\wpsi])_R = - \ii
  (u_i,\frac{\partial}{\partial \tau} \wpsi)_R = - \sum_j \omega_j (u_i,u_j)_R \war_j 
  = - \omega_i \war_i
$$
Likewise, the full set of commutation relations can be obtained:
\begin{align}\label{commrel2}
  [\wPi_R,\war_i] &= - \omega_i \war_i  &  [\wPi_R,\wadr_i] &= \omega_i \wadr_i 
 & [\wPi_R,\wal_i] &= [\wPi_R,\wadl_i] = 0 \nonumber \\
 [\wPi_L,\wal_i] & = \omega_i \wal_i  &  [\wPi_L,\wadl_i] &= - \omega_i \wadl_i 
 & [\wPi_L,\war_i] &= [\wPi_L,\wadr_i] = 0 
\end{align}
which show that $\wPi_R$ and $\wPi_L$ play the role of Hamiltonian operators. 

It is also worth pointing out that an explicit expression of the operators $\wPi_R, 
\wPi_L$ can can be obtained in terms of Rindler creation and destruction operators.
To show it, one extends the Klein-Gordon inner product (\ref{innprod}) to operators, 
by defining 
\footnote{In (\ref{innprod2}) it is understood that anytime the product of field 
operators appears, anticommutation is implied. For instance
$$
  \wpsi_1^\dagger \nabla^\mu \wpsi_2 \equiv \frac{1}{2} \{\wpsi_1^\dagger, 
   \nabla^\mu \wpsi_2\}
$$
so that the order of operators appearing in (\ref{innprod2}) does not matter 
even if they do not commute with each other.}:
\be\label{innprod2}
  (\wpsi_1,\wpsi_2) = i \int_\Sigma \di \Sigma_\mu \; \left( \wpsi_1^\dagger 
  \nabla^\mu \wpsi_2 - \wpsi_2 \nabla^\mu \wpsi_1^\dagger  \right)
\ee
. With this definition, it can be shown
that \cite{crispino}:
\be\label{generat2}
  \wPi = \wPi_R - \wPi_L = \frac{\ii}{2} \left( \wpsi,\gamma \cdot \nabla \wpsi 
  \right) = \frac{\ii}{2} \left( \wpsi, \frac{\partial}{\partial \tau} \wpsi \right)
\ee
Thus, taking the relations (\ref{orthon}) into account:
\begin{eqnarray*}
  \wPi &=& \frac{\ii}{2} \left( \wpsi, \frac{\partial\wpsi}{\partial\tau} \right) 
  \nonumber \\
 &=& \frac{\ii}{2} \sum_{i,j} (u_i \war_i + u^*_i \wadr_i, -\ii \omega_j u_j \war_j
 + \ii \omega u^*_j \wadr_j )_R + (u_i \wadl_i + u^*_i \wal_i, -\ii \omega_j u_j 
  \wadl_j + \ii \omega u^*_j \wal_j )_L \nonumber \\
 &=& \frac{1}{2} \sum_{i} \omega_i \left (\wadr_i \war_i + \war_i \wadr_i \right)-
  \omega_i \left (\wadl_i \wal_i + \wal_i \wadl_i \right) \nonumber \\
\end{eqnarray*}
and, by using the (\ref{commr}), (\ref{comml}):
\be\label{explicit}
  \wPi = \sum_{i} \omega_i \left (\wadr_i \war_i - \wadl_i \wal_i \right)
\ee
whence the commutation relations (\ref{commrel2}) could be derived. The last expression
shows that:
$$
 \wPi_{R} = \sum_{i} \omega_i \wadr_i \war_i \qquad \qquad
 \wPi_{L} = \sum_{i} \omega_i \wadl_i \wal_i
$$
From the eq.~(\ref{commrel2}) the following relations ensue:
\begin{align}\label{commrel3}
  \exp[\wPi_R/T_0] \war_i \exp[-\wPi_R/T_0] &= \e^{-\omega_i/T_0} \war_i  
& \exp[-\wPi_R/T_0] \wadr_i \exp[\wPi_R/T_0] &= \e^{-\omega_i/T_0} \wadr_i \nonumber \\
 \exp[-\wPi_L/T_0] \wal_i \exp[\wPi_L/T_0] &= \e^{-\omega_i/T_0} \wal_i  
& \exp[\wPi_L/T_0] \wadl_i \exp[-\wPi_L/T_0] &= \e^{-\omega_i/T_0} \wadl_i  
\end{align}
which are needed to calculate the mean values of products of creation and destruction 
operators. For instance, $\langle \wadr_i \war_j \rangle$ can be determined by using the 
eqs.~(\ref{commr}), (\ref{commrel3}) and taking advantage of trace ciclicity:
\begin{align}\label{ciclo}
 \langle \wadr_i \war_j \rangle &= \frac{1}{Z} \tr \left( \exp[-\wPi/T_0] \wadr_i  
 \war_j \right) = \e^{-\omega/T_0} \frac{1}{Z} \tr \left( \wadr_i \exp[-\wPi/T_0] 
 \war_j \right) \nonumber \\
 &= \e^{-\omega/T_0} \frac{1}{Z} \tr \left( \war_j \wadr_i \exp[-\wPi/T_0] \right) 
  = \e^{-\omega/T_0} \frac{1}{Z} \tr \left( \wadr_i \war_j \exp[-\wPi/T_0] \right)
  + \e^{-\omega/T_0} \delta_{ij} \nonumber \\
 & = \e^{-\omega/T_0} \langle \wadr_i \war_j \rangle + \e^{-\omega/T_0} \delta_{ij} 
\end{align}
The above equation has one finite solution, the well known Bose-Einstein distribution:
\be\label{bec}
  \langle \wadr_i \war_j \rangle = \delta_{ij} \frac{1}{\e^{\omega/T_0} - 1}
\ee
Similarly, it can be easily shown that:
\be\label{zero}
 \langle \wadr_i \wadr_j \rangle = \langle \war_i \war_j \rangle = 0 
\ee

Conversely, in the LRW, a similar derivation leads to:
\be\label{ciclo2}
 \langle \wadl_l \wal_j \rangle = \e^{\omega/T_0} \langle \wadl_i \wal_j \rangle
  + \e^{\omega/T_0} \delta_{ij}
\ee
which has a negative algebraic solution. A positive definite solution can be 
obtained by iteration, starting from:
$$ 
 \langle \wadl_i \wal_j \rangle_0 = \e^{\omega/T_0} \delta_{ij}
$$
replacing it on the right hand side of eq.~(\ref{ciclo2}) and iterating. Eventually:
$$
  \langle \wadl_i \wal_j \rangle = \delta_{ij} \sum_{k=1}^\infty \e^{k\omega/T_0}
$$
which is mainfestly a divergent series. That the mean occupation number in the LRW is infinite
is a clear consequence of the negative time component of the four-temperature therein.
Under this circumstance, the higher energy states with large number of quanta are 
{\em favoured}, unlike in the RRW. However, this does not hurt if one is to calculate 
the mean values of local operators in the RRW as this divergent factor in the trace
is cancelled by the partition function normalizing factor. 

We are now in a position to write the general expression of the mean value (\ref{generic})
in the RRW.
By using the (\ref{bec}) and (\ref{zero}) and using the commutation relation, we
are left with a single sum over the modes, like in the usual thermal quantum field 
theory:
\be\label{generic2}
  \langle A \wpsi B \wpsi \rangle = \int_0^{+\infty} \di \omega \int \di^2 {\rm k}_T \; 
 \left[ f_{\omega,{\bf k}_T}(x) \frac{1}{\e^{\omega/T_0} - 1} + f_{\omega,{\bf k}_T}^*(x) 
  \left(\frac{1}{\e^{\omega/T_0} - 1} + 1 \right) \right]
\ee
The complete calculation of physical interesting quantities which are quadratic in
the field, such as stress-energy tensor or currents, requires the specification of
the forms $A,B$. Once they are known, exact expressions for $f_{\omega,{\bf k}_T}$ can 
be obtained by using the field eigenfunctions (\ref{eigenf}). The terms in the
integrand of (\ref{generic2}) involving Bose-Einstein distributions give rise to 
a finite value, whereas the term arising from the $+1$ within brackets is divergent. 

As an example, we can reckon the mean value of the Lorentz-invariant expression, 
which enters in the calculation of the stress-energy tensor:
\be\label{udot}
  \langle u \cdot \partial \wpsi \; u \cdot \partial \wpsi \rangle = \frac{1}{\gamma^2}
  \langle \gamma \cdot \partial \wpsi \; \gamma \cdot \partial \wpsi \rangle = 
  \e^{-2a\xi} \big\langle \left( \frac{\di \wpsi}{\di \tau} \right)^2 \big\rangle 
\ee
where we have used the $\gamma$ definition (\ref{gammadef}), the eq.~(\ref{betacc2})
and the (\ref{rinverse}). In general, local Lorentz scalar functions such as (\ref{udot}) 
can depend on the Lorentz scalars $\beta^2$ and $A^2$; for instance, the canonical 
stress-energy tensor for the massless scalar field \cite{buzze} has a non-vanishing 
quadratic correction in $A^2$. By using the (\ref{eigenf}) it can be shown that
for the quantity (\ref{udot}) the functions $f_{\omega,{\bf k}_T}$ in eq.~(\ref{generic2})
are real and read:
\be
  f_{\omega,{\bf k}_T} = \omega^2 \frac {\sinh (\pi\omega/a)}{4 \pi^4 a}
  {\rm K}_{\ii \omega/a} \left(\frac{m_T \e^{a\xi}}{a} \right) 
  {\rm K}^*_{\ii \omega/a} \left(\frac{m_T \e^{a\xi}}{a} \right)
\ee
In the massless case $m_T = k_T$ and the integration can be done analitically:
$$
  \int \di^2 {\bf k}_T  {\rm K}_{\ii\omega/a} \left(\frac{m_T \e^{a\xi}}{a} \right) 
  {\rm K}_{-\ii\omega/a} \left( \frac{m_T \e^{a\xi}}{a} \right)
  = \frac{\pi a^2}{\e^{2a\xi}}\Gamma(1+\ii\omega/a) \Gamma(1-\ii\omega/a) 
  = \frac{\pi^2 a}{\e^{2a\xi}}\frac{\omega}{\sinh (\pi\omega/a)}
$$
where we have taken into account that ${\rm K}^*_{\ii\omega/a} = {\rm K}_{-\ii\omega/a}$.
Thus, for the non-divergent part of the eq.~(\ref{udot}) we have:
\be\label{expval}
\langle u \cdot \partial \wpsi \; u \cdot \partial \wpsi \rangle \rightarrow 
  \frac{1}{2 \pi^2} \e^{-4 a \xi} \int_0^{+\infty} \di \omega \; \omega^3 
  \frac{1}{\e^{\omega/T_0}-1} = \frac{\pi^2}{30} \frac{\e^{-4a\xi}}{T_0^4}
  = \frac{\pi^2}{30} \frac{1}{(\beta^2)^2}
\ee
This local mean value only depends on $\beta^2$.

\section{Subtraction of the vacuum expectation value}

The mean value of a local operator quadratic in the fields at equilibrium 
in eq.~(\ref{generic2}) features divergencies owing to the constant term $+1$ which
stems from the commutation relations of creation and destruction operators. In the 
familiar thermal free field theory with density operator (\ref{homo1}), the analogous
term is renormalized away by subtracting the vacuum expectation value. Yet, in the 
present case, we face an ambiguity as the vacuum state is dependent on which field 
expansion is taken. Since the Bogoliubov relations (\ref{bogo}) mix creation and 
destruction operators, the Rindler vacuum, i.e. the state which is annihilated by 
both the $\war$ and $\wal$ operators:
$$
  \war \ket{0_{\rm R}} =  \wal \ket{0_{\rm R}} = 0
$$
does not coincide with the usual Minkowski vacuum $\vacm$.

Which vacuum contribution to subtract then? In order to solve the ambiguity, we 
can argue that the subtraction of the Minkowski vacuum is better motivated, both 
physically and from a mathematical viewpoint. Indeed, the inertial observer is a 
privileged one in Minkowski space-time, and one should then refer to its vacuum. 
Furthermore, as 
it can be realized from the previous construction, the Rindler vacuum is a state 
explicitely dependent on the value of $a$,  which is a thermodynamic parameter appearing 
in the density operator (\ref{accdo3}). In fact, the vacuum state should not be
dependent on the density operator, and this is an undesired feature of the Rindler
vacuum which is not shared by the Minkowski vacuum. We then conclude that the most 
appropriate renormalization procedure is to subtract the Minkowski vacuum contribution, 
hence for a general quadratic operator:
\be\label{renquad}
  \langle A\wpsi B\wpsi \rangle_{\rm ren} = \langle A\wpsi B\wpsi \rangle
   - \vacmb A\wpsi B\wpsi \vacm
\ee

Unruh proved that the Minkowski vacuum expectation value corresponds to a thermal 
Bose-Einstein distribution of particles at a temperature $T_0 = a/(2\pi)$ \cite{unruh}. 
Even if it is a well known result, we believe that it is worth outlining a derivation 
based on an analytical prolongation method which is applicable to free as well as 
an interacting theory \cite{bisognano}. It should be pointed out that the rigorous 
mathematical proof involves many subtleties, what is presented below is a largely 
simplified version. 

In Minkowski spacetime, for a general Lorentz boost and a scalar field one has:
$$
  \exp[-\ii \xi \wK'_z] \wpsi(x) \exp[ \ii \xi \wK'_z] = 
  \wpsi( \e^{-\ii \xi {\sf K'}_z} x)
$$
where $\xi$ is the boost hyperbolic angle. This relation can be analitically extended
to imaginary $\xi$'s up to $\xi= \ii \pi$ \cite{bisognano,bisognano2}. For this limiting
value:
$$
  \exp[ \pi \wK'_z] \wpsi(t,x,y,z') \exp[ - \pi \wK'_z] = 
  \wpsi(-t,x,y,-z')
$$
so $\exp[\pi {\sf K'}_z]$ is in fact a rotation of $\pi$ in the $(t,z')$ plane. We
can now apply the above operators to the Minkowski vacuum $\vacm$. Since:
\be\label{kvacuum}
 \exp[ - \pi \wK'_z] \vacm = \exp[ - \pi (\wK_z - \widehat H /a )] \vacm
 = \exp( \pi E_{0}/a) \vacm
\ee
where $E_0$ is the vacuum energy and we have used the (\ref{kshift}) and the invariance 
of the vacuum under Lorentz transformations, that is $\wK_z \vacm = 0$. Hence
we get:
$$
 \exp( \pi E_{0}/a) \exp[ \pi \wK'_z] \wpsi(t,x,y,z') \vacm  = \wpsi(-t,x,y,-z') \vacm
$$
We now turn to the corresponding relations in the Rindler coordinates. First, we
note that the right hand side must be mapped onto the LRW, with the same coordinates 
$\tau$ and $\xi$ as on the left hand side (see eqs.~(\ref{rrindler}),(\ref{linverse})).
Hence, by plugging the expansions (\ref{rfield}),(\ref{lfield}), solving the Klein-
Gordon inner products (\ref{norm}),(\ref{norm2}) and using the (\ref{relkpi}) we 
obtain:
\be\label{vacuum1}
  \exp( \pi E_{0}/a) \exp \left[ - \frac{\pi}{a} \wPi \right] \war \vacm = \wadl \vacm \qquad 
  \exp( \pi E_{0}/a) \exp \left[ - \frac{\pi}{a} \wPi \right] \wadr \vacm = \wal \vacm \qquad
\ee
Now, by using the relations (\ref{commrel3}) and taking into account that, because
of the eqs.~(\ref{relkpi}) and (\ref{kvacuum}):
$$
  \exp[- \pi \wPi/a] \vacm = \exp[ \pi \wK'_z] \vacm = \exp( - \pi E_{0}/a) \vacm
$$
the equations (\ref{vacuum1}) turn into:
\be\label{vacuum2}
   \e^{\pi\omega_i/a} \war_i \vacm = \wadl_i \vacm \qquad 
   \e^{-\pi\omega_i/a~} \wadr_i \vacm = \wal_i \vacm 
\ee
These relations and the eqs.~(\ref{commr}),(\ref{comml}),(\ref{discon}) imply that 
\cite{crispino}:
\be\label{unruh}
  \vacmb \wadr_i \war_j \vacm = \vacmb \wadl_i \wal_j \vacm = \delta_{ij} 
  \frac{1}{\e^{2\pi \omega_i/a} -1}
\ee
which is the celebrated Unruh result.

We are now in a position to evaluate the Minkowski vacuum expectation value. This
is but the same expression as in eq.~(\ref{generic}) with $\vacmb \; \vacm$ replacing 
$\langle \; \rangle$:
$$
  \vacmb A\wpsi B\wpsi \vacm = \int_0^{+\infty} \di \omega \int \di^2 {\rm k}_T \; 
 \left[ f_{\omega,{\bf k}_T}(x) 
 \vacmb \wadr_{\omega,{\bf k}_T} \war_{\omega,{\bf k}_T} \vacm + f_{\omega,{\bf k}_T}(x)^*  
 \vacmb \war_{\omega,{\bf k}_T}\wadr_{\omega,{\bf k}_T} \vacm) \right]
$$
Hence, by using the above equation along with the (\ref{commr}), (\ref{generic2}) 
and the (\ref{unruh}), the eq.~(\ref{renquad}) becomes:
\be\label{main}
  \langle A\wpsi B\wpsi \rangle_{\rm ren} = \int_0^{+\infty} \di \omega \int \di^2 {\rm k}_T \; 
  (f_{\omega,{\bf k}_T}(x)+ f_{\omega,{\bf k}_T}(x)^*) \left( \frac{1}{\e^{\omega/T_0} - 1} - 
  \frac{1}{\e^{2 \pi \omega/a} - 1} \right)
\ee
The above expression is very suggestive. Any quadratic operator in the fields, 
including the stress-energy tensor, vanishes at $T_0 = a/(2\pi) \equiv T_U$. This 
may lead to different conclusions: if the functions $f_{\omega,{\bf k}_T}$ 
are positive definite, then the mean value turns negative when the temperature $T_0$ 
becomes lower than the Unruh temperature $T_{0U} = a/(2\pi)$. Alternatively, $T_{0U}$ 
is an absolute lower bound for the temperature $T_0$, where the mean values vanish.  
It is important to stress that this conclusion holds locally, for a comoving observer
or thermometer. Because of the relation (\ref{comovt}), the inequality $T_0 > a/(2\pi)$ 
implies:
\be\label{localbound}
  T = \frac{T_0}{a} \sqrt{-A^2} > \frac{\sqrt{-A^2}}{2 \pi} \equiv T_U
\ee
that is the temperature measured by a comoving thermometer cannot exceed the 
magnitude of the four-acceleration divided by $2\pi$, which can be rightly defined 
as the {\em comoving} Unruh temperature $T_U = |A|/(2\pi)$. 

This feature is apparent in the renormalized mean value of a scalar quantity, such 
as the one in eq.~(\ref{udot}). Applying the general expression (\ref{main}) to
the mean value in eq.~(\ref{udot}) and taking (\ref{expval}) into account we otbain:
$$
\langle u \cdot \partial \wpsi \; u \cdot \partial \wpsi \rangle_{\rm ren}=
\langle u \cdot \partial \wpsi \; u \cdot \partial \wpsi \rangle - 
\vacmb u \cdot \partial \wpsi \; u \cdot \partial \wpsi \vacm = 
\frac{\pi^2}{30} (T^4 - T^4_U)
$$
If the mean value at equilibrium with the density operator (\ref{accdo}) of a scalar 
quantity also depends on the magnitude of the acceleration (see refs.~\cite{becagrossi,
buzze} for a study of the stress-energy tensor with a perturbative expansion in 
$a/T_0$) this conclusion holds. Indeed, as has been mentioned, any mean local scalar 
quantity can be written as a function of the two scalars at our disposal, that is 
$T^2$, and $A^2/T^2$. In formula:
$$
 F\left( T^2,\frac{A^2}{T^2} \right) = F\left(\frac{T^2_0}{k^2 a^2},\frac{a^2}{T^2_0}
 \right)
$$
where we have used the eqs.~(\ref{tt0}) and (\ref{comovt}). By subtracting the term 
with $T_0=a/(2\pi)$ according to the eq.~(\ref{main}) one has:
$$
  F\left(\frac{T^2_0}{k^2 a^2},\frac{a^2}{T^2_0}\right) - F \left( \frac{1}{(2\pi)^2 k^2}, 
  (2\pi)^2 \right) = F\left( T^2,\frac{A^2}{T^2} \right)- F\left( T^2_U,\frac{A^2}{T^2_U} 
  \right)
$$
that is, for any local thermodynamic function we must subtract the corresponding 
value setting $T=T_U$.

\section{Discussion}

\begin{figure}[ht]
\includegraphics[width=0.8\columnwidth]{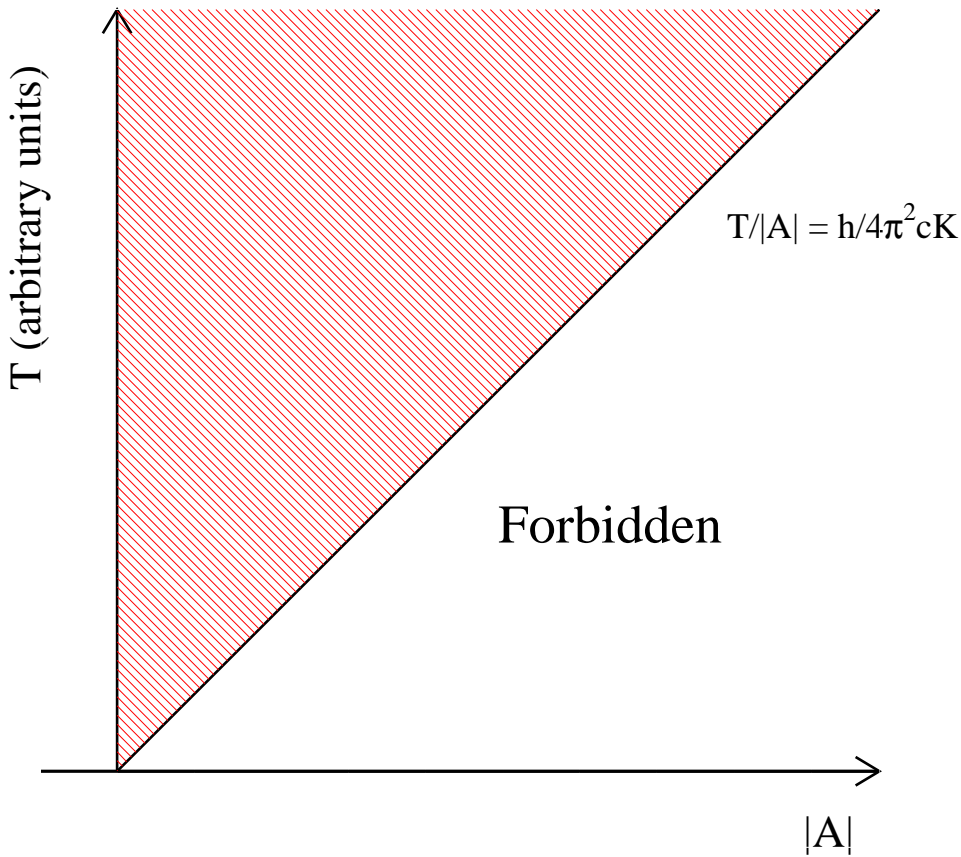}
\caption{(Color online) In the $T-|A|$ phase diagram of an accelerated fluid at 
global equilibrium there is a forbidden region delimited by the line $T = (h/4\pi^2 cK) 
|A|$.}
\label{phase}
\end{figure}

The usual phrasing of the Unruh effect is that an accelerated observer will measure
a thermal radiation bath while the inertial observer sees none in the Minkowksi 
vacuum state. In this work we have actually shown that the effect can be rephrased 
as follows: an ideal thermometer comoving in an accelerated fluid at thermodynamic 
equilibrium (that is, with the acceleration field pertaining to the global thermodynamic 
equilibrium situation described by the four-temperature Killing field in eq.~(\ref{betacc})) 
cannot measure a temperature lower than $|A|/(2\pi)$ (see fig.~\ref{phase}). This 
conclusion perfectly agrees with the above phrasing of the Unruh effect insofar 
as the accelerated comoving thermometer still sees a thermal radiation when the 
fluid stress-energy tensor vanishes in the Minkowski vacuum state. 
We stress that this effect is local: in the Killing flow (\ref{betacc}) the comoving
Unruh temperature depends on the magnitude of the acceleration, which is constant
along a single Killing flow line but varies for different flow lines \footnote{In 
this respect, our conclusion differs from the one in ref.~\cite{buchholz}.}

Although we have shown this for a free real scalar field, the argument can be 
extended. Particularly, the Unruh effect was rederived for an interacting scalar 
field theory in ref.~\cite{unruhweiss} where it was shown that: 
$$
 \vacmb {\rm T} [\wpsi(x),\ldots,\wpsi(x')] \vacm = \frac {\tr 
 (\exp[-2\pi\widehat K'/a] {\rm T} [\wpsi(x),\ldots,\wpsi(x')])}
 {\tr (\exp[-2\pi\widehat K'/a])}
$$
where ${\rm T}$ is the time-ordered product. This result implies, if $\wrho$ is 
the density operator in eq.~(\ref{accdo}):
\be\label{general}
  \langle \widehat O(x) \rangle_{\rm ren} = \tr (\wrho (T_0) \widehat O(x)) 
  - \vacmb \widehat O(x) \vacm = \tr (\wrho (T_0)\widehat O(x)) - 
  \tr((\wrho (a/(2\pi))\widehat O(x))
\ee
for any local operator. The eq.~(\ref{general}), as we have shown at the end of 
the previous section, entails that scalar thermodynamic functions like energy 
density or pressure, for an accelerated fluid at equilibrium can be written as:
\be\label{thermfunct}
   p(T^2,A^2) = p_{\rm th} (T^2,A^2) - p_{\rm th}(T_U^2, A^2)  
\ee
where $p_{\rm th}$ is the function calculated with the density operator (\ref{accdo}).

This conclusion is likely to hold for {\em any} interacting field theory, i.e. for 
any fluid. Indeed, the Unruh effect was derived for a general interacting field 
theories in refs.~\cite{bisognano,bisognano2} within axiomatic quantum field theory
approach taking advantage of the KMS feature of the mean values for the density 
operator at hand (for recent studies see refs.~\cite{pinamonti,gransee}).  

Finally, we briefly address the issue of an arbitrary motion of the fluid. 
As has been mentioned in Sect.~\ref{relstatmech}, a density operator like (\ref{accdo})
can be obtained from a local Taylor expansion of the four-temperature $\beta$ field, which
in local thermodynamic equilibrium is not constrained to be a Killing vector field.
Thus, the question arises whether the found lower bound for the local temperature holds.
Indeed, the local form of the bound (\ref{localbound}), which states that the comoving 
local temperature is limited by the comoving acceleration, seemingly suggest that the 
thermodynamic bound may apply to local thermodynamic equilibrium as well, provided
that the corresponding $a$ and $T_0$ obtained from the Taylor expansion of $\beta$
vary much more slowly in space and time compared to the microscopic lengths. However, 
one should take into account that the Unruh phenomenon arises because of a non-local 
property, which is the presence of an event or Killing horizon hypersurface where 
$\beta^2=0$. For an arbitrarily moving fluid, in general, there is no Killing horizon 
for the actual four-temperature vector field. Therefore, the existence of a lower 
bound for local temperature in hydrodynamics, as well as for general gravitational 
fields, remains an open issue.

\section*{Acknowledgments}

We acknowledge useful discussions with A. Cotrone and E. Grossi.



\end{document}